\documentstyle[preprint,aps]{revtex}
\draft
\begin{document}
\newcommand{\be}{\begin{equation}}
\newcommand{\ee}{\end{equation}}
\title{Periodic Orbits and Spectral Statistics of Pseudointegrable
Billiards}
\author{Debabrata Biswas\thanks{email: biswas@kaos.nbi.dk}
\thanks{on leave from Theoretical Physics Division,
Bhabha Atomic Research Centre, Bombay 400~085}}
\address{Center for Chaos and Turbulence Studies,
Niels Bohr Institute \\
Blegdamsvej 17, 
Copenhagen $\O$, Denmark}
\maketitle
\vskip 0.2 in
\begin{abstract}

\par We demonstrate for a generic pseudointegrable billiard that
the number of periodic orbit families with length less than $l$ increases
as $\pi b_0l^2/\langle a(l) \rangle$, where $b_0$ is a constant
and $\langle a(l) \rangle$ is the average area occupied by 
these families. We also find that $\langle a(l)
\rangle$ increases with $l$ before saturating. Finally, we 
show that periodic orbits provide a good estimate of spectral 
correlations in the corresponding quantum spectrum and thus
conclude that diffraction effects are not as significant
in such studies.

\end{abstract}

\vskip 1.5in

\par Billiards are an interesting and well studied class of 
hamiltonian systems that display a wide variety of dynamical
behaviour depending on the shape of the boundary. Of these
pseudointegrable billiards form a sub-class and these correspond
to rational angled polygonal shaped enclosures with a particle 
reflecting specularly from the walls \cite{PJ}.
They possess two constants of motion like their integrable
counterparts \cite{rem0} but,
the invariant surface is topologically equivalent to a sphere
with multiple holes \cite{PJ} and not a torus.
As an example, consider the billiard
in Fig. 1. For any trajectory,  
$p_x^2$ and $p_y^2$ are conserved. The
invariant surface consists of four sheets (copies) corresponding to the 
four possible momenta, ($\pm p_x, \pm p_y$) that it can have 
and the edges of these
sheets can be identified such that the resulting surface has the
topology of a double torus.

\par The classical dynamics  no longer has the simplicity of an
integrable system where a transformation to action and angle
co-ordinates enables one to solve the global evolution equations 
on a torus. On the other hand, the
dynamics is non-chaotic with the only interesting feature occurring
at the singular vertex with internal angle $3\pi/2$. Here,
families of parallel rays split and 
traverse different paths, a fact that limits the extent
of periodic orbit families. This is in contrast to integrable 
billiards (and to the $\pi/2$ internal angles in Fig. 1)
where families of rays do not see the vertex and continue
smoothly.  

\par We shall focus here on the periodic orbits of
such systems for they form the central object in modern 
semiclassical theories \cite{MC}. Not much is however known
about the manner in which they are organized and the 
few mathematical results that exist \cite{gutkin}
concern the asymptotic properties of their proliferation
rate. For a sub-class of rational polygons where the
vertices and edges lie on an integrable lattice (the
so called almost-integrable systems \cite{gutkin}), these
asymptotic results are exact. It is known for example 
that the number of periodic orbit families (those which
have even number of bounces), $N(l)$,  
increases quadratically with length, $l$, as $l \rightarrow 
\infty$. For general rational polygons, rigorous results
provide bounds on $N(l)$ though it is believed that  
$N(l) \sim l^2$ even in these cases \cite{gutkin,smillie}. 
Numerical verifications of these results are few and perhaps
none exist for a general 
rational polygon that does not belong to the category
of almost-integrable systems. Besides, very little is known
about other aspects such as the sum rules obeyed by periodic
orbits in contrast to the integrable and chaotic limits
where these have been well studied \cite{HO}.

\par It is not surprizing then
that periodic orbit theories for polygonal billiards have
met with little success, both in quantizing individual levels
\cite{richens83,shimizu}
and in explaining the correlations in the quantum spectrum
\cite {ccetal}. We shall not deal with the question of
determining individual levels here, but merely point
out, that, agreement with the smeared quantum density of
states has been observed using complex energies 
\cite{richens83,shimizu} although alternate convergent
schemes (such as cycle expansions in chaotic systems \cite{predrag}) 
are indeed desirable for better resolution of
individual levels. In this sense, the full scope of
periodic orbit quantization in such systems is largely
unknown, even though several researchers have now looked
beyond geometric periodic orbit contributions and
found evidence of diffractive corrections \cite{niall}.

\par Correlations on the other hand are related to  
sum rules obeyed by periodic orbits and these are
robust quantities that do not suffer from 
acute convergence problems. We thus focus on the
problem of computing 
correlations in the quantum spectrum using  
geometric periodic orbits. Further, our results
indicate that these estimates are good indicating
that diffraction effects are not as significant
in studies involving the statistical properties
of the spectrum \cite{niall2}. To appreciate this finding
however, it is necessary to study the sum
rules obeyed by periodic orbit families
and show that there are important differences
from the integrable case contrary to what
can be expected from the asymptotic properties
of $N(l)$ \cite{shimizu}. To this end, we first 
verify that periodic orbits in a generic pseudointegrable
system obey the sum rule \cite{prl1} $\langle \sum_p \sum_{r=1}
^{\infty} a_p\delta(l-rl_p)/(rl_p)\rangle = 2\pi b_0$, where $b_0$
is a constant, $\langle . \rangle$ denotes the average value,
$l_p$ is the length and $a_p$ the area of a 
primitive periodic orbit family. This establishes the
proliferation law, $N(l) = \pi b_0 l^2/ \langle a(l) 
\rangle$, derived in \cite{prl1} where 
$\langle a(l)\rangle $ is the average area occupied by
families having lengths $rl_p\leq l$. Further, we explore
the behaviour of $\langle a(l)\rangle$ as a function of length,
$l$, and find that it increases initially before saturating 
to a value much smaller than the maximum possible area spanned by a
single family.
The proliferation law, $N(l)$, is thus quadratic only asymptotically
and the number of orbits is much larger than that of an equivalent
integrable system with the same area. For smaller lengths, $N(l)$ 
is sub-quadratic and this is significant for a two-point correlation
of the quantum spectrum that we study and for which we 
demonstrate that periodic orbits
provide accurate estimates.

\par The L-shaped billiard of Fig.~(1) that we choose has 
approximately unit area and has no periodic orbit with
odd number of bounces at the boundary. It does not belong
to the class of almost-integrable billiards and is generic
in the sense that all sides are irrationally related and
periodic orbit lengths are non-degenerate. Unlike some
degenerate cases \cite{richens83} where periodic orbits 
can be labelled by two integers (analogous to the
winding numbers in 2-dimensional integrable systems), orbits
in generic L-shaped billiards are described by a 
set of four integers though not every point on the
$4$-dimensional lattice corresponds to a real periodic 
orbit. This makes it difficult to study sum rules from
purely classical considerations.     
An alternate approach adopted in \cite{prl1}
uses the semiclassical trace formula which expresses the
density of quantum energy eigenvalues in terms of periodic 
orbits \cite{PJ}: 

\be
 \sum_n \delta (E - E_n) = d_{av}(E) + {1\over 4\pi}
\sum_p \sum_{r=1}^{\infty} a_p J_0(krl_p) \label{eq:qtrace} 
\ee

\noindent
where $J_0$ is a Bessel function,
$\{E_n\}$ and $d_{av}(E)$ are the quantum energy eigenvalues
and their average density respectively,
and $k = \sqrt{E}$ \cite{neglect}.
For convenience, we have chosen the mass, $m = 1/2$ and $\hbar = 1$.
Starting with : 

\be
g(l) =  \sum_n f(\sqrt{E_n};\beta) =
\int_\epsilon^\infty dE f(E;\beta) \sum_n \delta(E-E_n) 
\ee

\noindent
where $f(\sqrt{E};\beta) = J_0(\sqrt{E}l)e^{-\beta E} $ and
$ 0 < \epsilon < E_0 $, it is possible to show using
eq.~(\ref{eq:qtrace}) that for $\beta \rightarrow 0^+$:

\be
\sum_p \sum_{r=1}^{\infty} {a_p \over rl_p} \delta (l-rl_p)
 = 2\pi b_0 + 2\pi \sum_n J_0(\sqrt{E_n}l) \label{eq:myown}
\ee

\noindent                   
where $b_0 = \sum_p \sum_{r=1}^{\infty} {a_p\over 4\pi}\int_0^{\epsilon} dE
J_0(\sqrt{E}l)J_0(\sqrt{E}rl_p) $.

\par It is argued in \cite{prl1} that $b_0$ is a constant
and we demonstrate this here by plotting 
$ S(l) = \sum_p\sum_r {a_p \over rl_p}$ in Fig.~2 where the summation is
restricted to all periodic orbits with $rl_p \leq l$. It follows
from Eq.~(\ref{eq:myown}) that $S(l) \simeq 2\pi b_0 l$.

\par Fig.~2 shows the behaviour of $S(l)$ for a rectangular
(integrable) and an L-shaped (pseudointegrable) billiard. 
In the former case, the orbit lengths, {$l_p$} can be expressed 
in terms of the winding numbers (M,N) on the torus
\cite{berry85,Pramana} 
while the areas \{$a_p$\} are four times the area, $A$, of the 
billiard except for bouncing ball orbits
for which they are twice the area.
For the pseudointegrable billiard, both the lengths and areas
are determined numerically using two different methods. We illustrate
one of these by first noting that one member of each periodic orbit 
family encounters the singular vertex. Further, a non-periodic orbit
originating
from the same point but with a momentum slightly different 
from a periodic orbit suffers a net transverse deviation 
that equals $(-1)^{n_\phi}\sin(\phi - \phi _p)l_{\phi}$. Here
$l_{\phi}$ is the distance traversed by a non-periodic 
orbit at an angle $\phi$ after $n_\phi$ reflections from the 
boundary and $\phi_p$ is the angle at
which a periodic orbit exists. These facts can be used to
converge on periodic orbits rapidly and the method works for other  
polygonal billiards. Details of this and the other method employed 
can be found in \cite{dbprep}. 

\par Note that in both cases, the curves in Fig.~2 are linear 
as expected from eq.~(\ref{eq:myown}). 
For the integrable case, $b_0 = 0.25$, while for the pseudointegrable
billiard $b_0 \simeq 0.27$ \cite{rem1}. This is the first 
difference between the two cases. The higher value for the
pseudointegrable billiard is possibly due to the fact that at 
the singular
vertex, there can exist more than one periodic orbit with the 
same value of $\phi_p$. It could also reflect diffraction 
effects that have been neglected in Eq.~(\ref{eq:qtrace}). These
issues 
will be discussed in a future publication \cite{dbprep}.

\par It is clear then that the leading term for the counting 
function, $N(l)$ is $\pi b_0l^2/\langle a(l) \rangle$
with corrections provided
by the the quantum energy eigenvalues. This result holds
for all rational polygons including those which are neither 
integrable nor almost-integrable. Here the average projected
phase space area, $ \langle a(l) \rangle \equiv (\sum_p \sum_r a_p)/N(l)$
where the summation extends over all orbits with $rl_p \leq l$.
For rectangular billiards, $\langle a(l) \rangle \simeq 4A$ and this gives the
quadratic law for $N(l)$ \cite{MC,Pramana}. For pseudointegrable
billiards, we plot $\langle a(l) \rangle$ in Fig. 3. The saturation for 
large, $l$ implies $N(l) \sim l^2$ asymptotically \cite{gutkin}.
For smaller values of $l$, $\langle a(l) \rangle$ increases,
indicating a (local) sub-quadratic law for $N(l)$ that we
have indeed verified. 

\par Note that the maximum area, $a_{max}$, occupied by a family is
four times the area of the billiard since the invariant surface
consists of four sheets corresponding to the four possible momenta a 
trajectory can have. The value at which $ \langle a(l) \rangle$
saturates is thus far smaller than $a_{max}$. Hence the density
of orbit lengths for the 
pseudointegrable billiard is far in excess of an equivalent integrable
billiard having the same area.

\par With these findings on periodic orbits, we now turn to the  
statistical analysis of the quantum spectra.
A commonly used measure is the spectral 
rigidity, $\Delta (L)$ defined as \cite{mehta}:

\be
\Delta (L) = \langle \min_{a,b}{d_{av}\over L} \int_{-L/2d_{av}}
^{L/2d_{av}} [N(E_0 + E) - a - bE]^2 dE \rangle \label{eq:delta3}
\ee

\noindent
where $N(E)$ counts the number of eigenvalues,
$E_0$ is the energy at which the measure is evaluated and 
$\langle . \rangle$ is 
an averaging in energy over scales larger than the outer scale
\cite{berry85} determined by the slowest frequency of oscillation
in Eq.~(\ref{eq:qtrace}). It is possible to analyze the rigidity in
terms of periodic orbits using eq.~(\ref{eq:qtrace}) and the 
basic semiclassical 
expression for the rigidity is then \cite{berry85} :

\be
\Delta (L) =  \langle \sum_i \sum_j {A_iA_j\over T_iT_j}
\cos(S_i - S_j) H_{ij} \rangle \label{eq:berdel} 
\ee

\noindent
where the summations extend over all periodic orbits \cite{rem2},
$A_i = (a_j^2/32\pi ^3 l_i\sqrt{E_0})^{1/2}$,$ T_i =
\partial S_i/\partial E $ evaluated at $E_0$ and 
$S_i = \sqrt{E_0}l_i$. The function $H_{ij} = F(y_i - y_j) - F(y_i)F(y_j) - 
3F'(y_i)F'(y_j)$ where  
$ y_i = LT_i/2d_{av}$ and $F(y) = \sin(y)/y$.

\par For $ L \ll 2\pi d_{av}/T_{min} $ and large $E_0$, 
Eq.~(\ref{eq:berdel})
can be simplified further to yield \cite{berry85} :

\be
\Delta(L) = {1\over 2\pi ^2}
 \int_0^{\infty} {d\tau \over \tau^2} K(\tau) G(\pi L \tau)
\label{eq:berdel1}
\ee

\noindent
where $G(y) = 1 - F^2(y) - 3F'^2(y)$, $\tau = T/(2\pi d_{av})$,
$K(\tau) = 2\pi \phi(T)/d_{av}$
and

\be
\phi (T) = \langle 
\sum_i \sum_j A_i A_j \cos(S_i - S_j) \delta(T - (T_i+T_j)/2) \rangle 
\ee

\par Eq.~(\ref{eq:berdel1}) is useful for analytical studies
only when the collective properties of periodic orbits as embodied
in $\phi(T)$ are known. A first step in this direction is the
diagonal approximation~\cite{berry85} for small $T$ based 
on the fact that orbit pairs have large action differences
and hence off-diagonal terms do not survive averaging. 
The diagonal sum, $\phi_D(T) = \langle \sum_i A_i^2\delta(T-T_i)
\rangle$ is however unknown for generic pseudointegrable
systems, though on treating the area, $a_i$, as constant 
and using the asymptotic law, $N(l) \sim l^2$, one might 
infer that $\phi_D(T)$ is constant \cite{shimizu} as in integrable 
systems \cite{berry85,HO}. These assumptions are however
incorrect especially for values of $T$ where the diagonal
approximation is expected to be valid. We shall therefore
investigate this numerically.

\par For large $T$, off-diagonal contributions are generally 
important and more difficult to estimate \cite{bogo}. For 
integrable
systems where the density of orbit lengths is small, off-diagonal
contributions vanish even for large $T$ \cite{berry85}. One might
expect this to be true even in the pseudointegrable situation
due to similarities in the asymptotic proliferation laws
though it must be noted that the density in the 
pseudointegrable case is
larger compared to an equivalent integrable system having
the same area.

\par In Fig.~4, we plot the function, $I(\tau) = \int_0^{\tau}
K(\tau ')d\tau '$ (upper curve) as well the diagonal part,
$I_D(\tau) = \int_0^{\tau} K_D(\tau ')d\tau '$.
The fact that $I_D(\tau)$ and $I(\tau)$ coincide
until $\tau_c \simeq 0.42$ implies that off-diagonal terms
do not contribute for $\tau < \tau_c$. The diagonal approximation
thus provides a good estimate of $K(\tau)$ for short times \cite{tauc}.

\par The important departure (as far as $\Delta(L)$ is concerned)
from integrable behaviour lies
in the fact that $I_D(\tau)$ displays variations in slope
for $\tau \leq \tau_c$ indicating that $K_D(\tau)$ is not
a constant \cite{seeprl1}. For $\tau > \tau_c$, $K_D(\tau)$ is constant
but has a value much smaller than unity (for integrable
systems $K_D(\tau) = 1$ for all values of $\tau$). Of 
significance as well is non-vanishing contribution of the off-diagonal
part ($I(\tau) - I_D(\tau)$) for $\tau > \tau_c$. This is 
probably the first example where a power law proliferation of
periodic orbits gives rise to non-zero off-diagonal contributions
in the form factor.

\par Finally, we compute the spectral rigidity using Eq.~(\ref{eq:berdel})
with 6621 periodic orbits and estimate the contribution of 
the longer ones using Eq.~(\ref{eq:berdel1}) and an interpolation for
$K(\tau )$. The result
is displayed in Fig.~5 where we also plot $\Delta(L)$ obtained
numerically using the quantum eigenvalues.
The agreement is remarkably good leading to the conclusion
that diffraction effects are not as significant for the 
statistical properties of the spectrum.

\par In summary, we have brought to light several interesting
properties of periodic orbits in pseudointegrable billiards
and numerically established that periodic orbits provide good
estimates of spectral correlations even when diffraction 
plays a role. Our specific conclusions are the following :

\par $\bullet$ orbit families obey the sum rule $\langle \sum_p \sum_{r=1}
^{\infty} a_p\delta(l-rl_p)/(rl_p)\rangle = 2\pi b_0$ thereby
giving rise to the proliferation law $N(l) = \pi b_0 l^2/ \langle
a(l) \rangle$ for all rational polygons.

\par $\bullet$ $\langle a(l) \rangle$ increases initially
before saturating to a value much smaller than $a_{max}$;
the asymptotic proliferation law is thus
quadratic even for systems that are not almost-integrable and
the density of periodic orbits lengths is far greater than
an equivalent integrable systems having the same area.

\par $\bullet$ the diagonal part of the form factor, $K(\tau)$
approaches a constant at $\tau = \tau_c$ and the value
is much smaller than unity. In contrast, the diagonal part
in integrable systems is identically equal to 1.

\par $\bullet$ off-diagonal contributions in $K(\tau)$ are non-zero 
for $\tau > \tau_c$ even though the proliferation of
periodic orbits is quadratic as in integrable billiards.

\par $\bullet$ periodic orbits provide good estimates of
correlations in the quantum spectrum. 

\nopagebreak
\vskip 0.25 in
\par It is a pleasure to acknowledge useful discussions with 
Predrag Cvitanovi\'{c}, Bertrand Georgeot, Gregor Tanner
and Niall Whelan.


\begin{thebibliography}{99}


\bibitem{PJ} P.J.Richens and M.V.Berry, Physica D{\bf 2}, 495(1981).
\bibitem{rem0} The few examples of integrable polygons are the
rectangle and triangles with internal angles ($\pi/3,\pi/3,\pi/3)$),
($\pi/3,\pi/6,\pi/2$) and ($\pi/4,\pi/4,\pi/2$).
\bibitem{MC} M.C.Gutzwiller, {\it Chaos in Classical and Quantum Mechanics},
Springer, New York, 1990.
\bibitem{gutkin} E.Gutkin, Physica D{\bf 19},311(1986).
\bibitem{smillie} See also W.Masur, Ergod. Th. and Dyn. Sys. {\bf 10},
151(1990).
\bibitem{HO} J.H.Hannay and A.M.Ozorio de Almeida,
J.Phys. A{\bf 17},3429(1984).
\bibitem{richens83} P.J.Richens, J.Phys. A{\bf 16}, 3961(1983).
\bibitem{shimizu} Y.Shimizu and A.Shudo, Chaos Solitons and Fractals,
{\bf 5}, 1337(1995).
\bibitem{ccetal} T.Cheon and T.D.Cohen, Phys. Rev. Lett. 
{\bf 62},2769(19891); A.Shudo, Y.Shimizu, P.Seba, J.Stein, H.Stockman
and K.Zyczkowski, Phys. Rev. E{\bf 49},3748(1994); D.Biswas and S.R.Jain,
Phys. Rev. A{\bf 42},3170(1990).
\bibitem{predrag} P.Cvitanovi\'{c}, Phys. Rev. Lett. {\bf 61}, 2729(1988). 
\bibitem{niall} N.D.Whelan, Phys. Rev. E {\bf 51},3778(1995);
N. Pavloff and C. Schmit, Phys. Rev. Lett. {\bf 75},61(1995).
\bibitem{niall2} Indirect evidence of this for the cardioid billiard
has been recently provided by H.Bruss and N.D.Whelan, Periodic
Orbit Theory of Edge Diffraction, chao-dyn/9509005. 
\bibitem{prl1} D.Biswas and S.Sinha, Phys. Rev. Lett. {\bf 70}, 916(1993).
\bibitem{neglect} We have neglected the effect of diffractive 
orbits in writing Eq.~(\ref{eq:qtrace}). The corrections are
$O(k^{-1 - \nu/2})$ where $\nu$ counts the number of (singular) vertex 
connections in a diffractive periodic orbit \cite{niall,niall2,richens83}.
\bibitem{berry85} M.V.Berry, Proc. Roy. Soc. Lond. Ser. A{\bf 400},
229(1985).
\bibitem{Pramana} D.Biswas, Pramana - Journal of Physics, {\bf 42},
447 (1994).
\bibitem{dbprep} D.Biswas, in preparation. 
\bibitem{rem1} For integrable billiards, the equality of the diagonal
and semiclassical sum rule for $\phi (T)$ that we discuss
shortly, implies $b_0 = 1/N$ \cite{prl1}. Here $N$
is the number of copies that are glued to form the 
torus. Thus for rectangular billiards, $b_0 = 1/4$ while for the 
equilateral triangle $b_0 = 1/6$. For reasons that can be found
in \cite{dbprep}, we expect $b_0 \simeq 1/4$ for the L-shaped
billiard as well.
\bibitem{mehta} M.L.Mehta, {\it Random Matrices }, Academic Press, 1967.
\bibitem{rem2} The index, $i$, treats repetitions as distinct periodic
orbits. 
\bibitem{bogo} see for example E.B.Bogomolny and J.Keating, Nonlinearity
{\bf 8}, 1115(1995) and references therein.   
\bibitem{tauc} The value of $\tau_c$ corresponds to the length $l$
at which the proliferation law becomes quadratic. 
\bibitem{seeprl1} Consequences of a local power law behaviour in
$K_D(\tau)$ can be found in \cite{prl1}.

\end{thebibliography}
\end{document}